\pdfoutput=1

\documentclass[11pt,twoside,a4paper,cmspaper,final,collab]{cms-tdr}
\def\svnVersion{488109P}\def\cmsCernNoTag{CERN-EP-2018-110}\def\cmsCernDate{\today}\def\cmsMessage{Published in Physics Letters B as \href{http://dx.doi.org/10.1016/j.physletb.2019.01.006}{\doi{10.1016/j.physletb.2019.01.006.}}}

\begin{document}\cmsNoteHeader{HIN-16-023}

\hyphenation{had-ron-i-za-tion}
\hyphenation{cal-or-i-me-ter}
\hyphenation{de-vices}
\RCS$HeadURL: svn+ssh://svn.cern.ch/reps/tdr2/papers/HIN-16-023/trunk/HIN-16-023.tex $
\RCS$Id: HIN-16-023.tex 488021 2019-02-01 00:59:58Z jaebeom $

\newlength\cmsTabSkip\setlength{\cmsTabSkip}{1ex}
\newlength\cmsFigWidth
\ifthenelse{\boolean{cms@external}}{\setlength\cmsFigWidth{0.49\textwidth}}{\setlength\cmsFigWidth{0.6\textwidth}}
\ifthenelse{\boolean{cms@external}}{\providecommand{\cmsLeft}{top\xspace}}{\providecommand{\cmsLeft}{left\xspace}}
\ifthenelse{\boolean{cms@external}}{\providecommand{\cmsRight}{bottom\xspace}}{\providecommand{\cmsRight}{right\xspace}}\ifthenelse{\boolean{cms@external}}{\providecommand{\CL}{C.L.\xspace}}{\providecommand{\CL}{CL\xspace}}
\ifthenelse{\boolean{cms@external}}{\providecommand{\appLabel}{}}{\providecommand{\appLabel}{Appendix}}

\cmsNoteHeader{HIN-16-023}

\newcommand{\Pb}{\ensuremath{\mathrm{Pb}}\xspace}
\newcommand{\pp}{\ensuremath{\Pp\Pp}\xspace}
\newcommand {\PbPb}{\ensuremath{\Pb\Pb}\xspace}
\newcommand {\AAa}{\ensuremath{\mathrm{AA}}\xspace}
\newcommand {\pPb}{\ensuremath{\Pp\Pb}\xspace}

\newcommand{\RAA}{\ensuremath{R_\AAa}\xspace}
\newcommand{\TAA}{\ensuremath{T_{\AAa}}\xspace}
\newcommand{\Tc}{\ensuremath{T_\mathrm{c}}\xspace}
\newcommand{\NMB}{\ensuremath{N_{\mathrm{MB}}}\xspace}
\newcommand{\Npart}{\ensuremath{N_{\text{part}}}\xspace}
\newcommand{\Ncoll}{\ensuremath{N_{\text{coll}}}\xspace}

\title{Measurement of nuclear modification factors of $\PgUa$, $\PgUb$, and $\PgUc$ mesons in \PbPb collisions at $\sqrtsNN = 5.02\TeV$}

\date{\today}

\abstract{
The cross sections for $\PgU$(1S), $\PgU$(2S), and $\PgU$(3S) production in lead-lead (\PbPb) and proton-proton
(\pp) collisions at $\sqrtsNN = 5.02\TeV$  have been measured using the CMS detector at the LHC.
The nuclear modification factors, \RAA, derived from the \PbPb-to-\pp ratio of yields for each state,
are studied as functions of meson rapidity and transverse momentum, as well as \PbPb collision centrality.
The yields of all three states are found to be significantly suppressed, and compatible with a sequential ordering of the suppression, $\RAA(\PgU\mathrm{(1S)}) > \RAA(\PgU\mathrm{(2S)}) > \RAA(\PgU\mathrm{(3S)})$. The suppression of $\PgU$(1S) is larger than that seen at $\sqrtsNN = 2.76\TeV$, although the two are compatible within uncertainties. The upper limit on the \RAA of \PgUc\ integrated over \pt, rapidity and centrality is 0.096 at 95\% confidence level, which is the strongest suppression observed for a quarkonium state in heavy ion collisions to date.
}

\hypersetup{%
pdfauthor={CMS Collaboration},
pdftitle={Measurement of nuclear modification factors of Y(1S), Y(2S), and Y(3S) mesons in PbPb collisions at sqrtsNN = 5.02TeV},
pdfsubject={CMS},%
pdfkeywords={CMS, physics, bottomonium, quarkonium suppression, quark gluon plasma, heavy ion collisions}
}

\maketitle

\pagenumbering{arabic}

\section{Introduction}
\label{sec:intro}

The measurement of quarkonium production in heavy ion collisions is one
of the most promising ways to study the properties of strongly interacting matter at high energy density and temperature.
It has been predicted that in such an environment, a strongly interacting medium of deconfined quarks and gluons (the quark-gluon plasma, QGP) is formed~\cite{Karsch:2003jg, Shuryak:1977ut}. Bottomonium states have been the subject of studies in heavy ion collisions for several reasons. Bottomonia are produced during the early stages of collisions via hard parton scattering. Their spectral functions are modified as a consequence of Debye screening of the heavy-quark potential at finite temperatures~\cite{Matsui:1986dk, Digal:2001ue}, as well as by thermal broadening of their widths due to interactions with gluons~\cite{Brambilla:2008cx, Brambilla:2010vq}. These in-medium effects have been studied in numerical simulations of quantum chromodynamics (QCD) on a space-time lattice, and captured as real and imaginary components of the heavy-quark potential~\cite{Brambilla:2010cs}. One of the most remarkable signatures of these interactions with the medium is the sequential suppression of quarkonium states in heavy ion collisions compared to the production in proton-proton (\pp) collisions, both in the charmonium (\PJGy, \Pgy, $\chi_\cPqc$, etc.) and the bottomonium (\PgUa, \PgUb, \PgUc, $\chi_\cPqb$, etc.) families~\cite{Andronic:2015wma}. This scenario follows from the expectation that the suppression of quarkonia is stronger for states with smaller binding energy. The quarkonium yield can also increase in the presence of QGP, from the recombination of uncorrelated quarks~\cite{Krouppa:2016jcl, Brambilla:2016wgg, Emerick:2011xu, Du:2017qkv}. However, recombination-like processes for bottomonia are expected to be negligible compared to the charmonium family~\cite{Thews:2000rj,Gorenstein:2000ck,Andronic:2007bi}, because these processes are driven by the number of heavy-quark pairs present in a single event, which is much smaller for beauty than for charm. The dissociation temperatures for the \PgU\ states, above which suppression occurs, are expected to be correlated with their binding energies, and are predicted to be $T_{\text{dissoc}}\approx2\Tc$, 1.2\Tc and 1\Tc for the \PgUa, \PgUb, and \PgUc\ states, respectively, where \Tc is the critical temperature for deconfinement~\cite{Mocsy:2007xu}. Therefore, measurements of the yields of each \PgU\ state can provide information about the thermal properties of the medium during its hot early phase.

Modifications of particle production in nucleus-nucleus (\AAa) collisions are quantified using the nuclear modification factor, \RAA, which is the ratio of the yield measured in \AAa to that in \pp collisions, scaled by the mean number of binary NN collisions. Comparisons of the bottomonium data with dynamical models incorporating the heavy-quark potential effects found in high-temperature lattice QCD are thus expected to extend our understanding of the nature of colour deconfinement in heavy ion collisions. Measurements of both the charmonium (\PJGy\ and \Pgy)~\cite{Khachatryan:2016ypw,Khachatryan:2014bva,Adam:2015isa,Adam:2015rba} and bottomonium (\PgUa, \PgUb, and \PgUc)~\cite{Khachatryan:2016xxp, Abelev:2014nua} families have been carried out at a nucleon-nucleon (NN) center-of-mass energy of $\sqrtsNN = 2.76\TeV$ and, most recently, at $\sqrtsNN = 0.2\TeV$ at RHIC~\cite{Adare:2006ns,Adamczyk:2016srz,Adamczyk:2013poh}. At $\sqrtsNN = 5.02\TeV$, measurements by the CMS Collaboration show strong suppression of \JPsi\ and \Pgy\ mesons~\cite{CMS:2016ajg,Sirunyan:2017isk}, as well as of both \PgUb\ and \PgUc\ relative to the \PgUa\ ground state~\cite{CMS:2016ayg}. The suppression of the excited \PgUb\  relative to the  \PgUa\ ground state persists at very forward rapidity, $2.5<y<4$~\cite{Acharya:2018mni}. These measurements provide new constraints for theoretical models of the medium~\cite{Krouppa:2016jcl,Emerick:2011xu}.

In this Letter, we report measurements of the differential cross sections and nuclear modification factors for \PgUa, \PgUb, and \PgUc\ mesons using their decay into two oppositely charged muons in lead-lead (\PbPb) and \pp collisions at $\sqrtsNN = 5.02\TeV$. Results are presented as functions of the \PgU\ transverse momentum (\pt) and rapidity ($y$), as well as \PbPb collision centrality (\ie, the degree of overlap of the two lead nuclei). The data were collected with the CMS detector at the CERN LHC in 2015.

\section{The CMS detector}
\label{sec:detector}

The central feature of the CMS apparatus is a superconducting solenoid of 6\unit{m} internal diameter,
providing a magnetic field of 3.8\unit{T}. Within the solenoid volume are a silicon
 tracker, a lead tungstate crystal electromagnetic calorimeter, and a brass and scintillator
 hadron calorimeter, each composed of a barrel and two endcap sections. Muons are detected in the pseudorapidity interval of $\abs{\eta} < 2.4$ using gas-ionization chambers made
of three technologies: drift tubes, cathode strip chambers, and resistive-plate chambers. These are embedded
 in the steel flux-return yoke of the solenoid.
The silicon tracker is composed of pixel detectors followed by microstrip detectors. The \pt of muons matched to tracks reconstructed in the silicon detector is measured with a resolution between 1\% and 2\% for typical muons used in this analysis~\cite{Khachatryan:2010xn}. In addition, CMS has extensive forward calorimetry, including two steel and quartz-fiber Cherenkov  hadron forward (HF) calorimeters that cover the range of $2.9 < \abs{\eta} < 5.2$. The HF calorimeters are segmented into towers and the granularity is $\Delta\eta\times\Delta\phi =0.175 \times 0.175$ radians. These are used in the present analysis to select \PbPb collision events and to define their centrality class. Centrality, defined as the fraction of the total inelastic hadronic cross section with 0\% representing collisions with the largest overlap of the two nuclei, is determined experimentally using the total energy in both HF calorimeters~\cite{CMS:2011za}. A more detailed description of the CMS detector, together with a definition of the coordinate system and the kinematic variables, can be found in Ref.~\cite{Chatrchyan:2008zzk}.

\section{Data selection and simulation samples}
\label{sec:dimuSel}

The \PgU\ mesons are identified using their dimuon decay channel.
In both \pp and \PbPb collisions, the dimuon events are selected by a fast
hardware-based trigger system, which requires two muon candidates
in a given bunch crossing with no explicit requirement on the muon momentum beyond the intrinsic selection due to the acceptance coverage of the CMS muon detectors. In \pp collisions, this trigger registered an integral luminosity of 28.0\pbinv. The \PbPb data were taken with two triggers based on the same algorithm used for \pp data. The first mode, designed to enhance the event count for muon pairs from peripheral events, added an additional selection that the collision centrality be in the 30--100\% range. This trigger sampled the full integrated luminosity of 464\mubinv. The second mode, using just the \pp trigger alone, was prescaled during part of the data taking and therefore sampled a smaller effective integrated luminosity of 368\mubinv. Data taken with this latter trigger were used to analyze the yields in the 0--30\% and 0--100\% centrality bins.

In order to keep  hadronic collisions and reject beam-related background processes (beam-gas collisions and beam scraping events), an offline event selection is applied. Events are required to have at least one reconstructed primary vertex. In \pp collisions at least 25\% of the tracks have to pass a tight track-quality selection~\cite{1748-0221-9-10-P10009}. A filter on the compatibility of the silicon pixel detector cluster width and the vertex position is also applied~\cite{CMS:2010za}. The \PbPb collision events have an additional requirement of the presence of at least three towers in the HF on both sides of the interaction point with an energy above 3\GeV. The combined efficiency for this event selection, and the remaining contamination due to non-hadronic ultra-peripheral events which can raise the efficiency above 100\%, is ($99\pm2)\%$~\cite{Chatrchyan:2011sx,Khachatryan:2016odn}. The minimum-bias trigger requirement removes a negligible fraction of the events with a hard collision needed to produce \PgU\ mesons. We also studied a possible contamination from photoproduction processes in the peripheral region and found it to be negligible. Multiple-collision events (pileup) have a negligible effect on the measurement, since the average number of additional collisions per bunch crossing is approximately 0.9 for \pp and much smaller for \PbPb data.

Muons are selected in the kinematic range of $\pt^{\mu} > 4$\GeV and $\abs{\eta^{\mu}}<2.4$, and are also required to be reconstructed using the combined information of the tracker and muon detectors (so-called "global muons" defined in Ref.~\cite{Khachatryan:2010xn}). To remove cosmic ray muons, the distance of the muon track from the closest primary vertex must be less than 20\unit{cm} in the beam direction and 3\unit{mm} in the transverse direction. Pairs of oppositely charged muons are fitted with a common vertex constraint and kept if the fit $\chi^2$ probability is larger than 1\%. The studied dimuon kinematic range is limited to $\pt^{\MM} < 30\GeV$ and $\abs{y^{\MM}}<2.4$. Dimuons in this \pt range comprise 99\% of those passing all of the analysis selection criteria.

Simulated Monte Carlo (MC) \PgU\ events are used to calculate correction factors for all of the results presented, including the geometrical acceptance and reconstruction efficiency, as well as the trigger and offline selection efficiency. The samples are generated using \PYTHIA8.209 \cite{Sjostrand:2015za} for the \pp collisions and \PYTHIA8.209 embedded in \HYDJET1.9 for the \PbPb events~\cite{Lokhtin:2006hh}. The \PbPb simulation is tuned to reproduce the observed charged-particle multiplicity and \pt spectrum in \PbPb
data. The CMS detector response is simulated using \GEANTfour~\cite{Agostinelli:2002hh}. Since the simulated \pt spectrum of \PgU\ is not identical to the spectrum observed in data, an event-by-event weight is applied to the simulations in order to match the two distributions. The weight is given by a function fit to the ratio of data over MC \pt spectra.

\section{Analysis procedure}
\subsection{Signal extraction}
\label{sec:sigExt}
The yields of \PgU\ mesons are extracted using unbinned maximum-likelihood fits to the invariant mass spectra, following the same procedure for \pp and \PbPb data.
The signal of each \PgU\ state is modeled by a double Crystal-Ball (CB) function which is the sum of two CB functions~\cite{SLAC-R-236}. This choice together with leaving the width parameter for the first CB free in the fit, is made in order to account for the different mass resolution in the barrel compared to the endcap region of the detector. A parameter relates the widths of the two CB functions, the second one being constrained to be narrower. The mass and the two radiative-tail parameters of both CB functions for a given state are kept the same, as these are not affected by the detector resolution. The mass parameter of the ground state is left free to allow for possible shifts in the absolute momentum calibration of the reconstructed tracks. For the excited states (\PgUb\ and \PgUc), the yields can vary while all other fit parameters are fixed to be identical to those for the ground state except for the mean and width which are fixed to values found by multiplying those for \PgUa\  by the ratio of the published masses of the states~\cite{PDG2016}. In the \pp data fits, the two radiative-tail parameters and the parameter for the ratio of the two widths are allowed to vary within a Gaussian probability density function (PDF). The mean and the width of the constraining Gaussian function represent the average and its uncertainty, respectively, from the fits in all the rapidity bins of the analysis with no fixed parameters. In the \PbPb fits, in addition, the parameter for the fraction of the two CB functions is also constrained. In this case, the mean and the width of the constrained parameters represent the corresponding parameter values and their uncertainties from the \pp fits for each kinematic region. The background PDF is an error function multiplied by an exponential, with the yield, the error function's two parameters, and the decay parameter of the exponential all allowed to vary in the final fit. For bins with $\pt>6$\GeV, an exponential without the error function provides the best fit, and was used for the nominal result.

\begin{figure}[hbt!]
  \includegraphics[width=  0.45\textwidth]{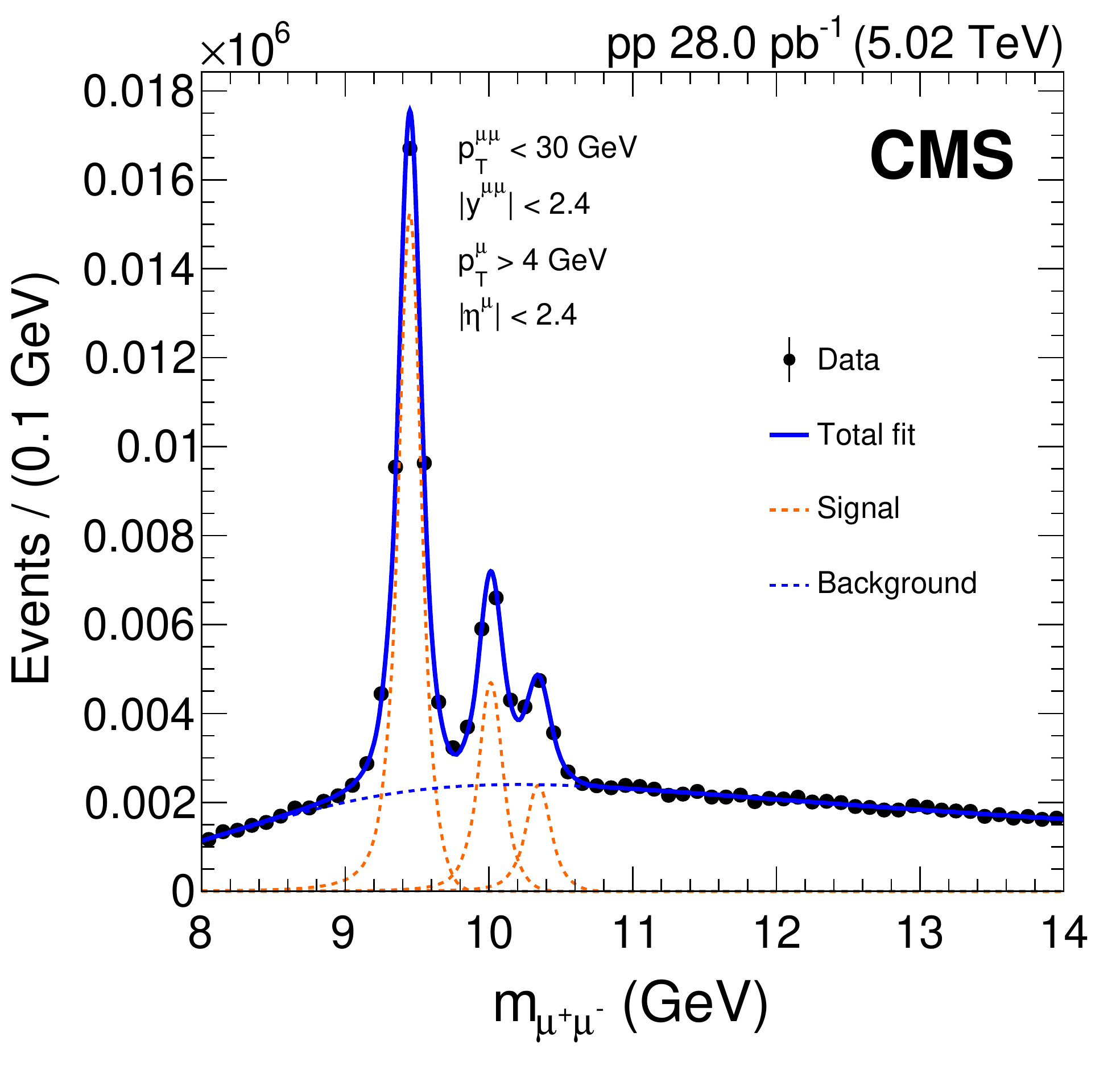}
  \includegraphics[width=  0.45\textwidth]{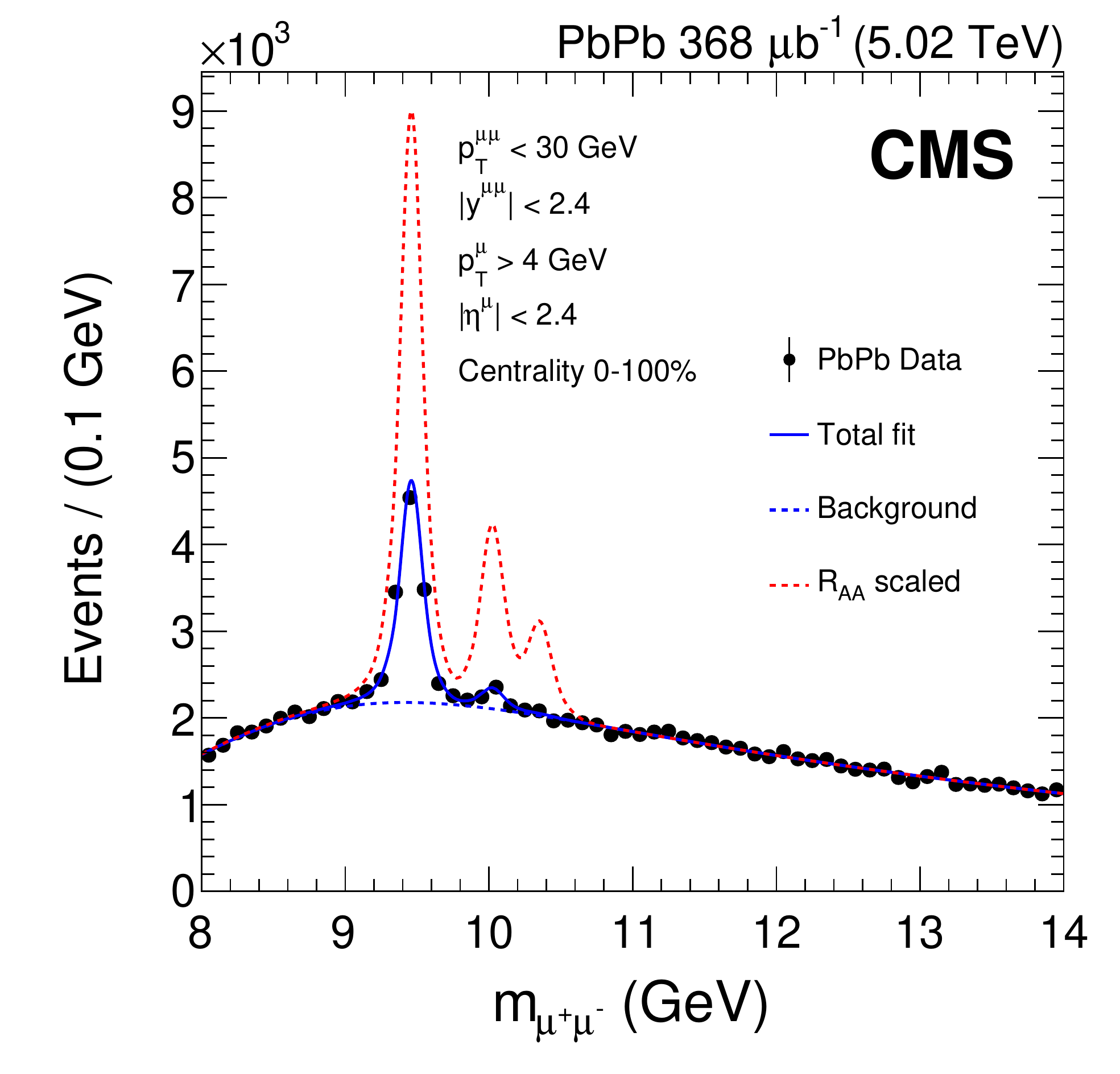}
  \caption{Invariant mass distribution of muon pairs in \pp (\cmsLeft) and \PbPb (\cmsRight) collisions, for the kinematic range $\pt^{\MM} < 30$\GeV and $\abs{y^{\MM}}<2.4$.
  In both figures, the results of the fits to the data are shown as solid blue lines. The separate yields for each \PgU\ state in \pp are shown as dashed red lines in the \cmsLeft panel. The dashed red lines in the \cmsRight panel are derived from the fits to \PbPb (blue solid line). 
In order to show the suppression of all three \PgU\ states, the amplitudes of the corresponding peaks are increased above those found in the fit by the inverse of the measured \RAA\ for the corresponding \PgU\ meson.}
  \label{fig:invMassFit}
\end{figure}

Figure~\ref{fig:invMassFit} shows the dimuon invariant mass distributions in \pp and \PbPb collisions along with the fits using the model described above, for the kinematic range $\pt^{\MM} < 30$\GeV and $\abs{y^{\MM}}<2.4$.

\subsection{Corrections}
\label{sec:corr}

In order to obtain the normalized cross sections, the yields extracted from the fits to the dimuon invariant mass spectra are corrected for acceptance and efficiency, and scaled by the integrated luminosity. The acceptance corresponds to the fraction of dimuon events originating from \PgU\ mesons within the kinematic range of the analysis. The acceptance values for the considered kinematic region are 22.5\% (\PgUa), 27.8\% (\PgUb), and 31.0\% (\PgUc) for \PbPb collisions and differ by $<$1\% from the corresponding \pp data values, with the small difference being due to a small residual difference in the kinematic spectra after weighting the MC to data.

The dimuon efficiency is defined as the probability that a muon pair within the acceptance is reconstructed offline, satisfies the trigger condition, and passes the analysis quality criteria described in Section~\ref{sec:dimuSel}. The dimuon efficiency is calculated using MC. The individual components of the efficiency (track reconstruction, muon identification and selection, and triggering) are also measured using single muons from \PJGy\ meson decays in both simulated and collision data, with the \textit{tag-and-probe} (T\&P) method~\cite{Khachatryan:2010xn}. For the muons used in this analysis, data and MC efficiencies are seen to differ only in the case of the trigger efficiency, and there only by $\la$1\%. For this case, scaling factors (SF), calculated as the ratio of data over simulated efficiencies as function of $\pt^{\mu}$ and $\eta^{\mu}$, are applied to each dimuon on an event-by-event basis. The other components of the T\&P efficiency are used only for the estimation of systematic uncertainties. The average efficiencies integrated over the full kinematic range are 73.5\% (\PgUa), 74.4\% (\PgUb), and 75.0\% (\PgUc) in \PbPb collisions, and they are 8--9\% higher for \pp collisions.

The integrated luminosity of 28.0\pbinv with an uncertainty of 2.3\%~\cite{CMS-PAS-LUM-16-001} is used to normalize the yields for \pp data. For \PbPb collisions, the number of minimum bias collision events sampled by the trigger (\NMB), together with the average nuclear overlap function (\TAA), are used for the normalization. The overlap function \TAA
is given by the number of binary NN collisions divided by the inelastic NN cross section, and can be interpreted as the NN-equivalent integrated luminosity per heavy ion collision. Values of \TAA are calculated with a Glauber model MC simulation~\cite{Miller:2007ri,Loizides:2014vua}, which is also used to obtain the average number of participating nucleons, $\langle \Npart \rangle$. This latter number is highly correlated with the impact parameter of the collision, and is used as the abscissa when plotting results as a function of \PbPb collision centrality.

\subsection{Systematic uncertainties}
\label{sec:sys}
Point-to-point systematic uncertainties arise from the choices of signal and background PDFs and of the central value in the fit constraints, as well as from acceptance and efficiency corrections. Larger relative uncertainties are obtained when the background level is higher (at lower \pt or more forward $y$ regions), and, in particular for the \PgUc, when the absolute yield is small.

The uncertainty from the choice of signal model is estimated by fitting the data using a single CB function in combination with a Gaussian function instead of a double CB function. The uncertainties are determined by calculating the difference between the yield obtained with the alternative model compared to the nominal one. For the \PbPb (\pp) yields, the differences are in the range of 1--7\% (0.1--4.6\%) for the \PgUa, 2--19\% (0.1--1.3\%) for the \PgUb, and 5--78\% (0.7--7\%) for the \PgUc\ mesons.

The systematic uncertainty from the choice of the central value in the fit constraints is estimated by using instead of the average parameter values from the \pp fits, the values in each \pp analysis bin when all parameters were left floating. The differences in the \PbPb (\pp) signal yields, typically below 4\% (4.5\%) for the \PgUa, below 8\% (3\%) for the \PgUb, and 45\% (2\%) for the \PgUc, are quoted as a systematic uncertainty.

The systematic uncertainty due to the choice of background model is estimated using two alternative background functions.
One is in the form of a fourth-order polynomial function and the other is an exponential plus an additional linear function.
The maximal deviations of the \PbPb (\pp) yield between these two models compared to the nominal are quoted as the uncertainty and are typically in the range of 1--6\% (1--5\%) for the \PgUa, 2--23\% (2--4\%) for the \PgUb, and 5--200\% (3--5\%) for the \PgUc\ mesons.

For the estimation of systematic uncertainties due to acceptance and efficiency corrections, the source of uncertainty is the imperfect knowledge of the simulated \pt distribution shape. To take this source into account, the function used to weight the MC \pt spectra event-by-event is modified within its fit uncertainty.
The acceptance and efficiency obtained from the simulated \pt distribution are compared with and without the variation of the function, with the difference between the two used as an estimate of the systematic uncertainty. In addition, there is a systematic uncertainty for the efficiency in the T\&P correction arising from the uncertainty in the SFs of the single-muon efficiency. The systematic uncertainties of the SFs are taken into account for trigger, tracking, and muon identification. The uncertainties in the single muon efficiencies are propagated to the dimuon efficiency values to estimate the systematic uncertainty from this source. The statistical uncertainty inherent in the data set used for the T\&P studies is also considered as an additional component of the systematic uncertainty in the corrected yields. The \PbPb (\pp) systematic uncertainties are in the range of 3.5--6.4\% (2.6--3.9\%) in the case of the total efficiency correction and in the range of 0.1--3.0\% (0.1--0.8\%) for the acceptance correction.

Finally, several sources of correlated uncertainties (\ie, global uncertainties common to all points) are considered: for the \pp dataset from the \pp integrated luminosity, and for the \PbPb dataset from the \TAA and the \NMB estimations.
The uncertainty on the integrated luminosity measurement for the \pp dataset is 2.3\%~\cite{CMS-PAS-LUM-16-001}.
The uncertainty for \NMB in \PbPb collisions is 2\%, which accounts for the inefficiency of trigger and event selection. For the \RAA calculation, \TAA uncertainties (Table~\ref{tab:CentralityNpart} in \appLabel~\ref{app:Supplement}) are estimated by varying the Glauber model parameters within their uncertainties (Table~\ref{tab:GlauberParameters} in \appLabel~\ref{app:Supplement})~\cite{Khachatryan:2016odn}. 
The total combined uncertainty is calculated by adding the results from the various sources in quadrature.
The global uncertainty for the differential cross section results arises from the integrated luminosity in \pp collisions and \NMB in \PbPb collisions. For the \RAA results, the global uncertainty combines the uncertainties from \TAA, \pp luminosity, and \PbPb \NMB for the bins integrated over centrality. For the centrality dependent \RAA results, the uncertainty from \TAA is included bin-by-bin, while the total uncertainty from the \pp measurement is included in the global uncertainty. Using the updated uncertainties of the Glauber model parameters in Ref.~\cite{PhysRevC.97.054910}, instead of those from Ref.~\cite{Khachatryan:2016odn}, would reduce the \TAA uncertainties by 0.1--1.1\% and the total systematic uncertainties for \RAA by less than 0.7\% (with the largest change for the 70--100\% centrality bin). However, in order to allow direct comparisons to previous results~\cite{Khachatryan:2016odn,Sirunyan:2017oug,Sirunyan:2017xss,Sirunyan:2017isk}, these updated parameters are not used in this analysis.   
The bin migration effect due to the momentum resolution is negligible for the kinematic range of this measurement.

\section{Results}
\label{sec:results}

The \PgU\ cross sections and values of \RAA are measured in several \pt and y bins. The rapidity studies are performed in the range $0< \abs{y} <2.4$. This rapidity range is evenly divided into six, three, and two bins for \PgUa, \PgUb, and \PgUc,  respectively. For the investigation of the behavior of the \RAA as a function of centrality, the bin limits of the centrality classes are chosen as follows: [0, 5, 10, 20, 30, 40, 50, 60, 70, 100\%] for the \PgUa\ and \PgUb, and [0, 30, 100\%] for the \PgUc. When plotted as a function of each variable (\pt, y or centrality), values are integrated over the full kinematic range of the other variables. The \PgUc\ mesons show a very strong suppression in \PbPb collisions, with yields which are statistically consistent with zero for all bins. The upper limits at 68\% and 95\% confidence level (\CL) for the \PgUc\ cross section and \RAA are found using the Feldman--Cousins method~\cite{Feldman:1997qc}, with the appropriate systematic uncertainties being included in the upper limit computation.

\subsection{Differential cross sections in \pp and \PbPb collisions}
\label{sec:CSresult}

The differential production cross section of \PgU\ mesons decaying in the dimuon channel in \pp collisions is given by
\begin{linenomath}
\begin{equation}
  \mathcal{B} \frac{\rd\sigma^2}{\rd y \rd\pt } = \frac{N/(\mathcal{A}\,\varepsilon)}{\mathcal{L}_{\text{int}}\Delta y \Delta\pt}.
\label{eqn:crosssection}
\end{equation}
\end{linenomath}
The branching fraction for the decay $\PgU \to \MM$ is denoted by $\mathcal{B}$. The quantity $N$ corresponds to the extracted yield of \PgU\ mesons in a given (\pt, $y$) bin, $(\mathcal{A}\,\varepsilon)$ represents the average acceptance and efficiency in the given bin, $\mathcal{L}_{\text{int}}$ is the integrated luminosity, and $\Delta \pt$ and $\Delta y$ are the widths of the given bin. For \PbPb data, $\mathcal{L}_{\mathrm{int}}$ is replaced by $(\NMB\,\TAA)$, as explained in Section~\ref{sec:corr}, to compare the \pp and \PbPb data under the hypothesis of binary-collision scaling.

\begin{figure}[hbtp]
\centering
\includegraphics[width=0.49\textwidth]{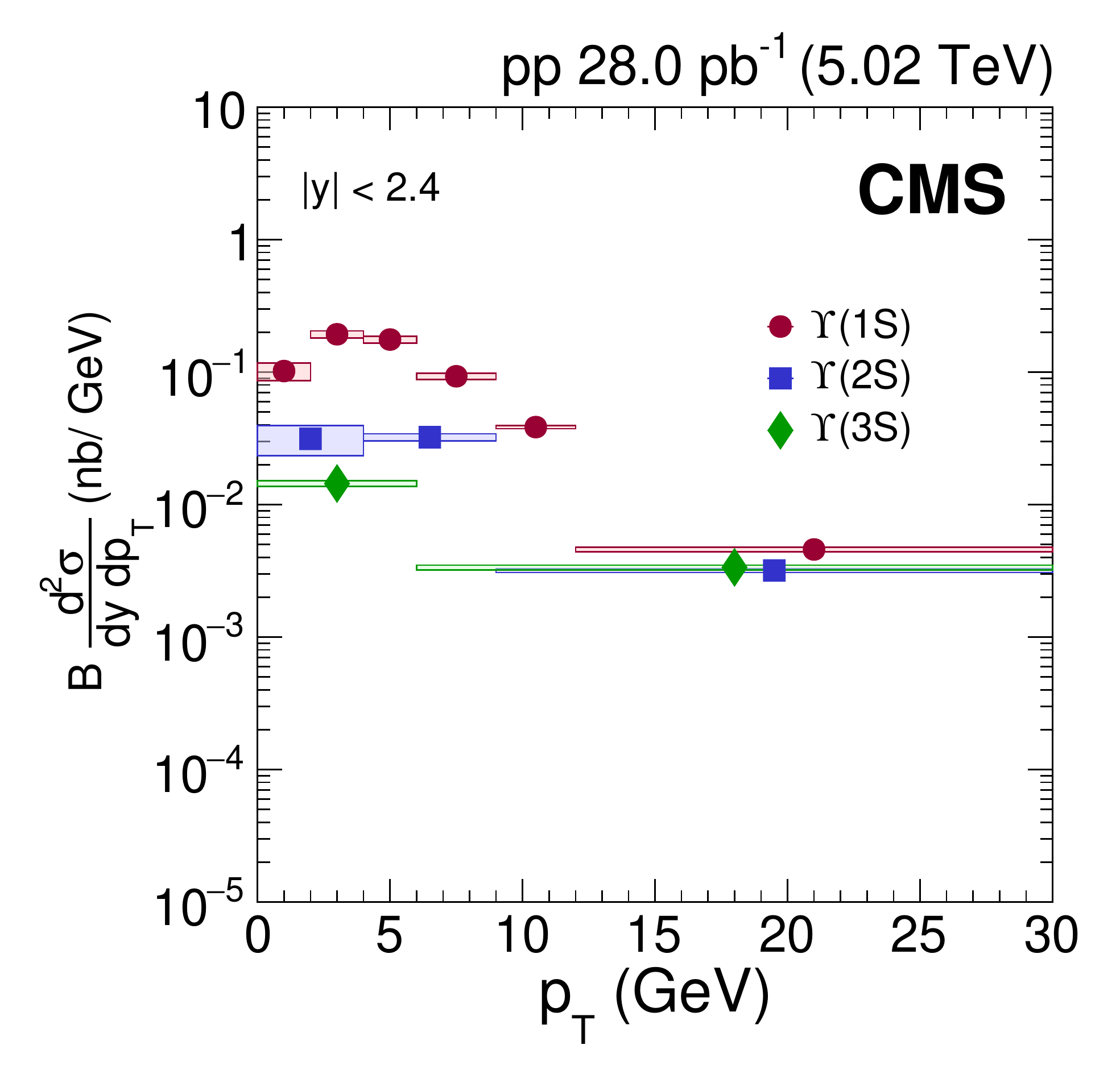}
\includegraphics[width=0.49\textwidth]{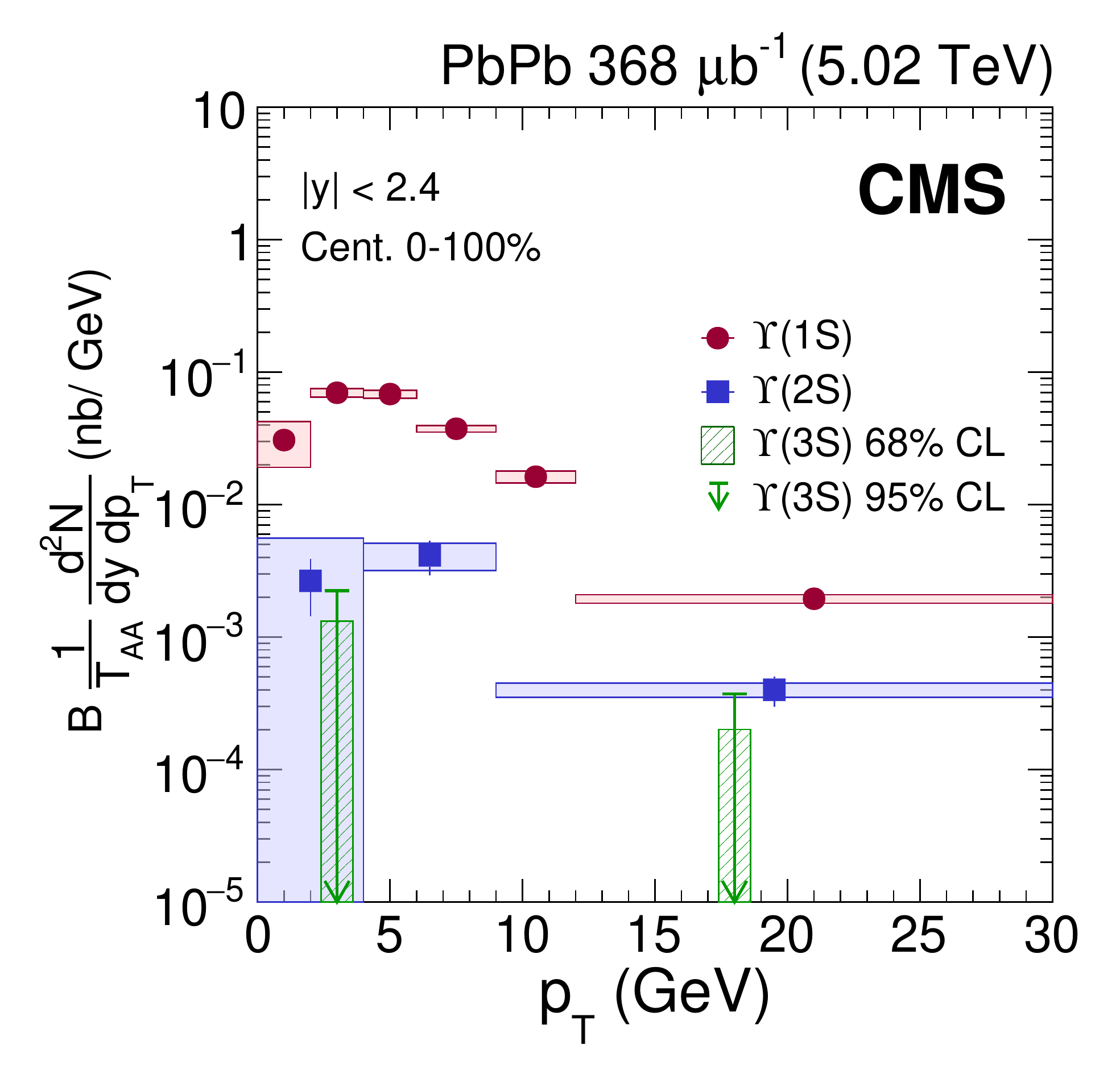}
\caption{\label{Fig:cs_pt} Differential cross sections of the \PgUa, \PgUb, and \PgUc\ mesons as a function of \pt for \pp (\cmsLeft) and \PbPb (\cmsRight) collisions.
The error bars represent the statistical uncertainties and the boxes the systematic uncertainties.
  For the \PgUc\ meson in \PbPb collisions, the upper limits at 68\% (green box)
  and 95\% (green arrow) \CL are shown, and are calculated for the same bins as for the \pp dataset.
    The global integrated luminosity uncertainties of 2.3\% in \pp collisions and $^{+3.4\%}_{-3.9\%}$ in \PbPb collisions are not shown.}
\end{figure}

\begin{figure}[hbtp]
\centering
\includegraphics[width=0.49\textwidth]{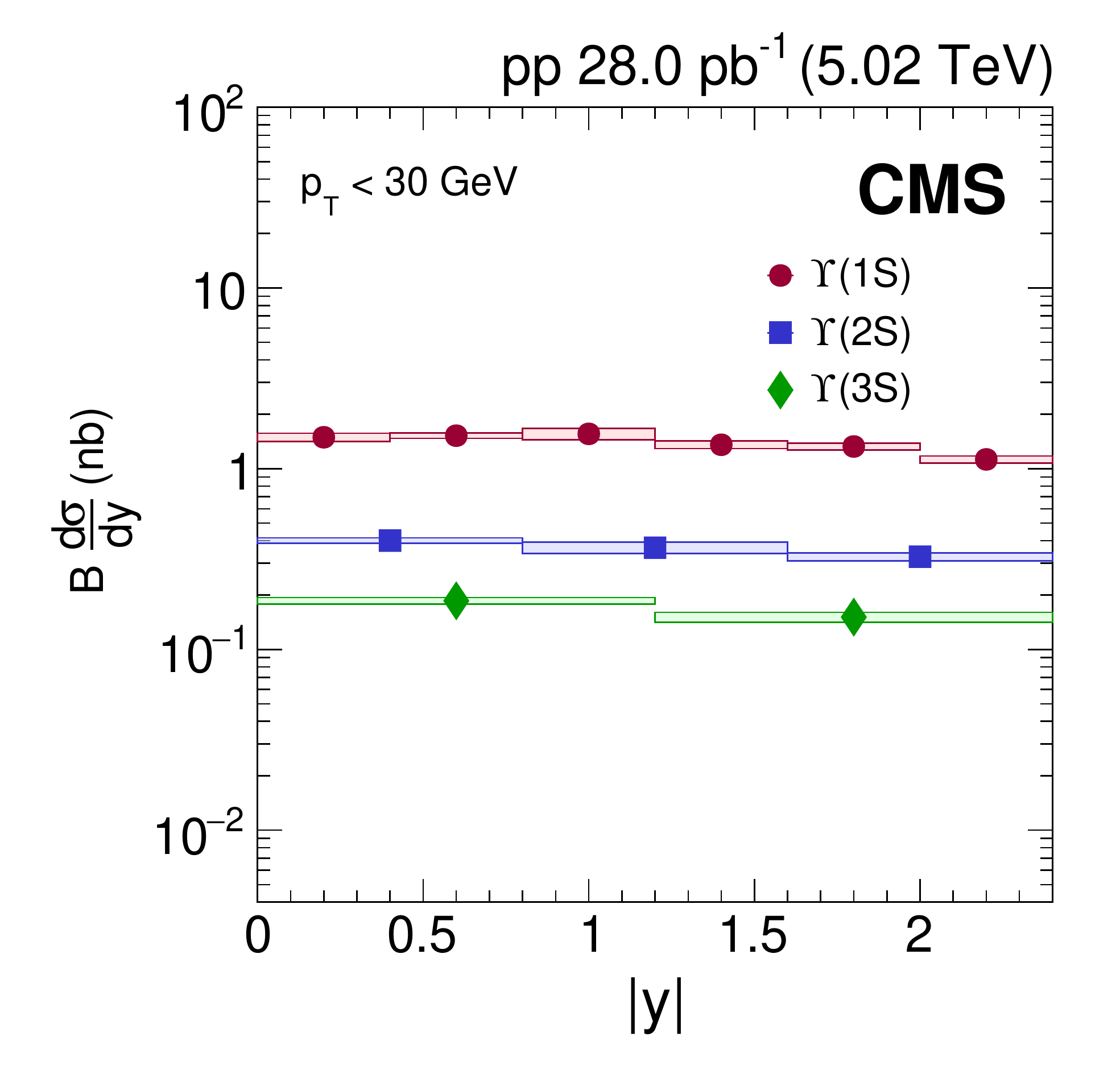}
\includegraphics[width=0.49\textwidth]{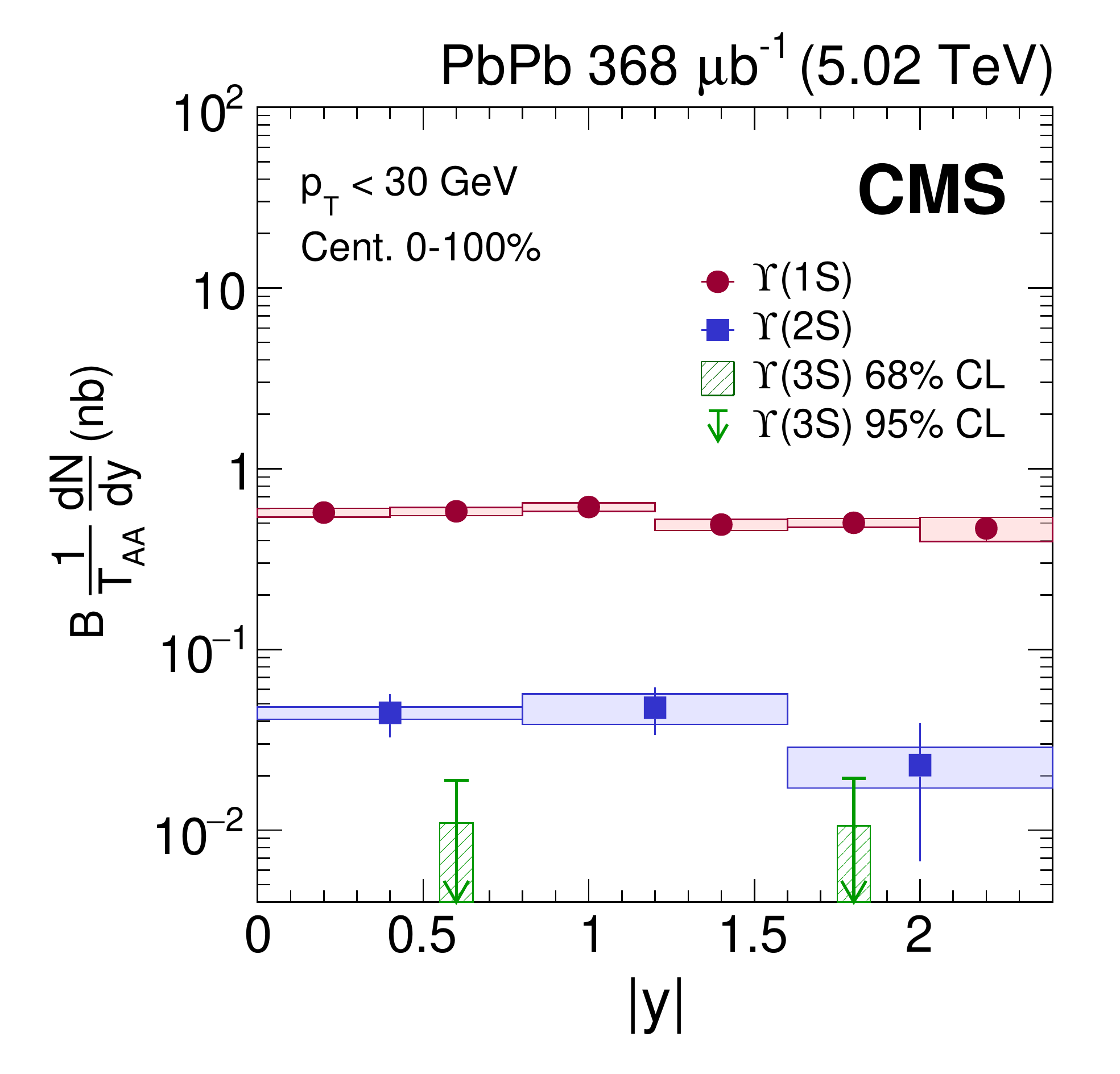}
\caption{\label{Fig:cs_rap} Differential cross sections of the \PgUa, \PgUb, and \PgUc\ mesons as a function of rapidity for \pp (\cmsLeft) and \PbPb (\cmsRight) collisions. The error bars represent the statistical uncertainties and the boxes the systematic uncertainties. For the \PgUc\ meson in \PbPb collisions, the upper limits at 68\% (green box) and 95\% (green arrow) \CL are shown. The global integrated luminosity uncertainties of 2.3\% in \pp collisions and $^{+3.4\%}_{-3.9\%}$ in \PbPb collisions are not shown.}
\end{figure}

Figure~\ref{Fig:cs_pt} shows the differential production cross sections of \PgU\ mesons as a function of \pt in \pp and \PbPb collisions. The data points are placed at the center of each bin. The corresponding results as a function of $\abs{y}$  are shown in Fig.~\ref{Fig:cs_rap}.

\subsection{Nuclear modification factor \texorpdfstring{\ensuremath{\RAA}}{R\_AA}}
\label{sec:RAAresult}

The nuclear modification factor is derived from the \pp cross sections and \PbPb normalized yields as
\begin{linenomath}
\begin{equation}
\RAA (\pt,y) = \frac{ N^{\textrm{AA}}(\pt,y) }{ \langle \TAA \rangle \sigma^{\pp}(\pt,y) },
\end{equation}
\end{linenomath}
where $\langle \TAA \rangle$ is the average value of \TAA computed in each centrality bin. The quantities $N^{\AAa}$ and $\sigma^{\pp}$ refer to the normalized yield of \PgU\ mesons in \PbPb collisions corrected by acceptance and efficiency, and the \pp cross section for a given kinematic range, respectively.

\begin{figure}[hbt!]
\centering
\includegraphics[width=0.48\textwidth]{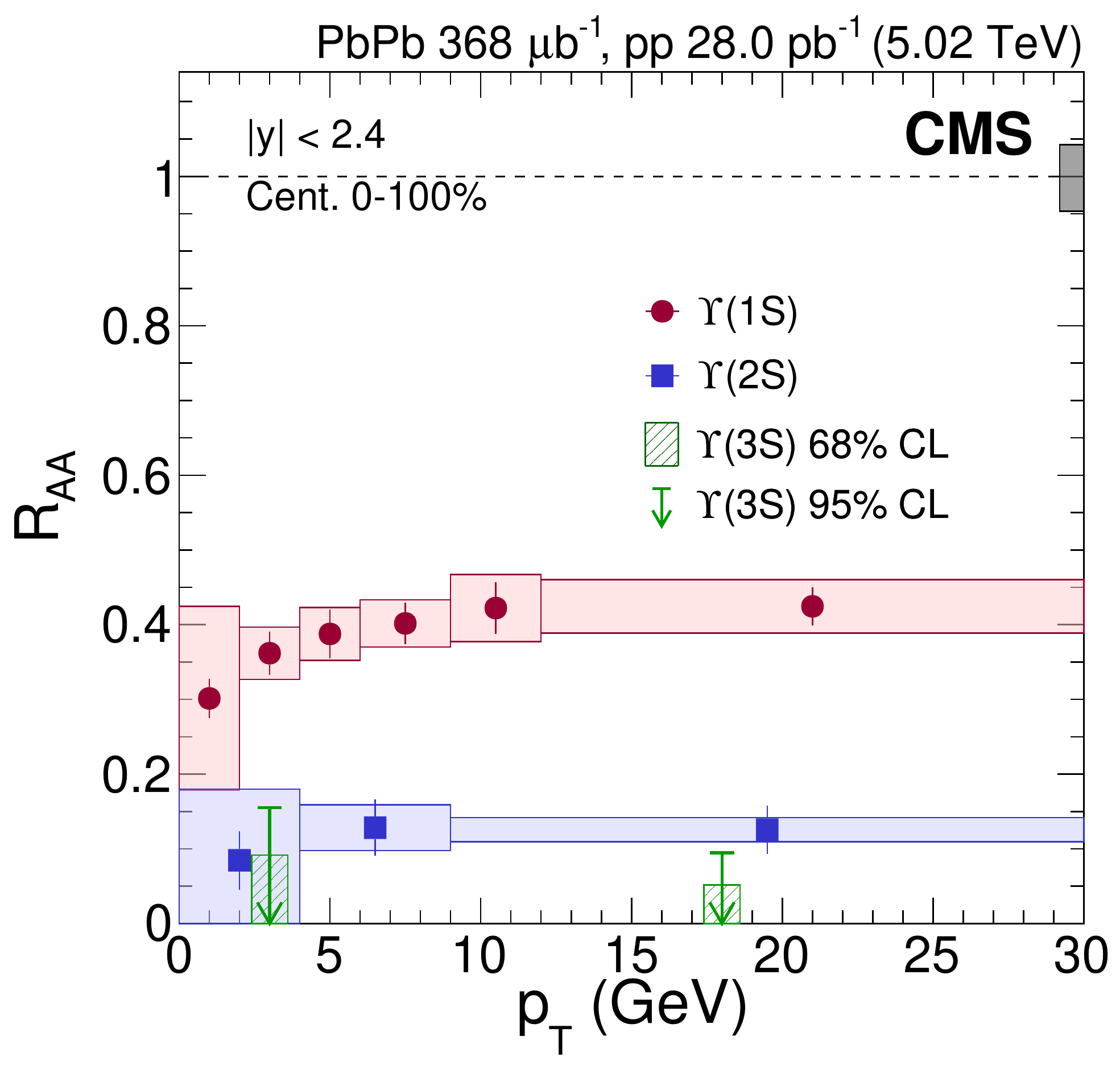}
\includegraphics[width=0.48\textwidth]{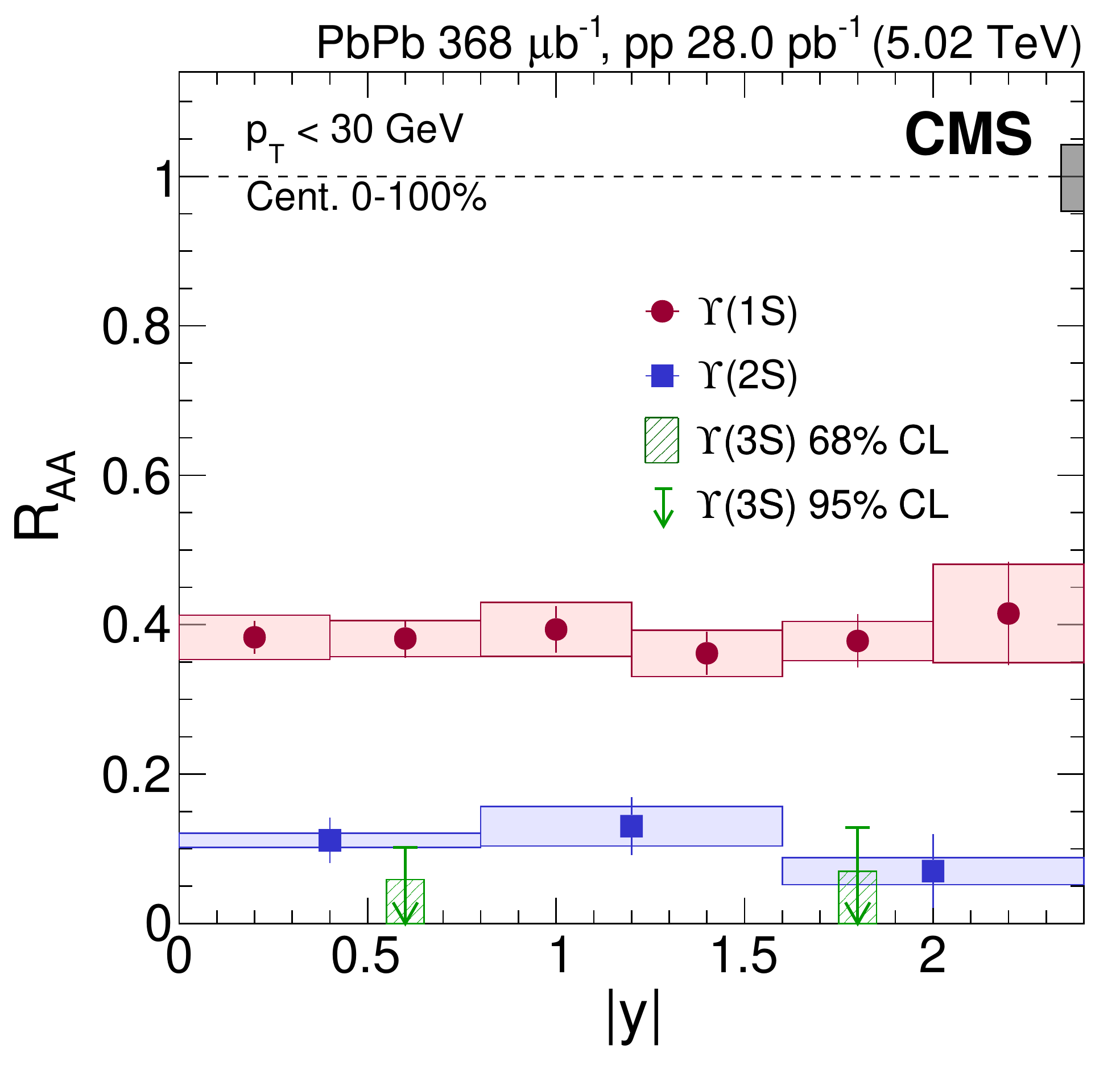}
\caption{\label{Fig:raa_rap_pt} Nuclear modification factors for \PgUa, \PgUb, and \PgUc\ mesons as functions of \pt (\cmsLeft) and rapidity (\cmsRight).
The error bars represent the statistical uncertainties and the boxes the systematic uncertainties.
For the \PgUc\ meson, the upper limits at 68\% (green box) and 95\% (green arrow) \CL are shown.
The gray box near the line at unity displays the global uncertainty, which combines the uncertainties from \TAA, \pp luminosity, and \PbPb \NMB. }
\end{figure}

Figure~\ref{Fig:raa_rap_pt} shows the nuclear modification factor for the \PgUa, \PgUb, and \PgUc\ mesons as functions of \pt and $\abs{y}$. Within the systematic uncertainties, the \RAA values show no clear dependence on \pt or $y$. The excited \PgU\ states are found to have larger suppression than the ground state, with $\RAA < 0.2$ over the full kinematic range explored here. The kinematic dependence of \RAA is useful to constrain models of \PgU\ meson suppression in a deconfined medium \cite{Krouppa:2016jcl}.

\begin{figure}[hbt!]
\centering
\includegraphics[width=\cmsFigWidth]{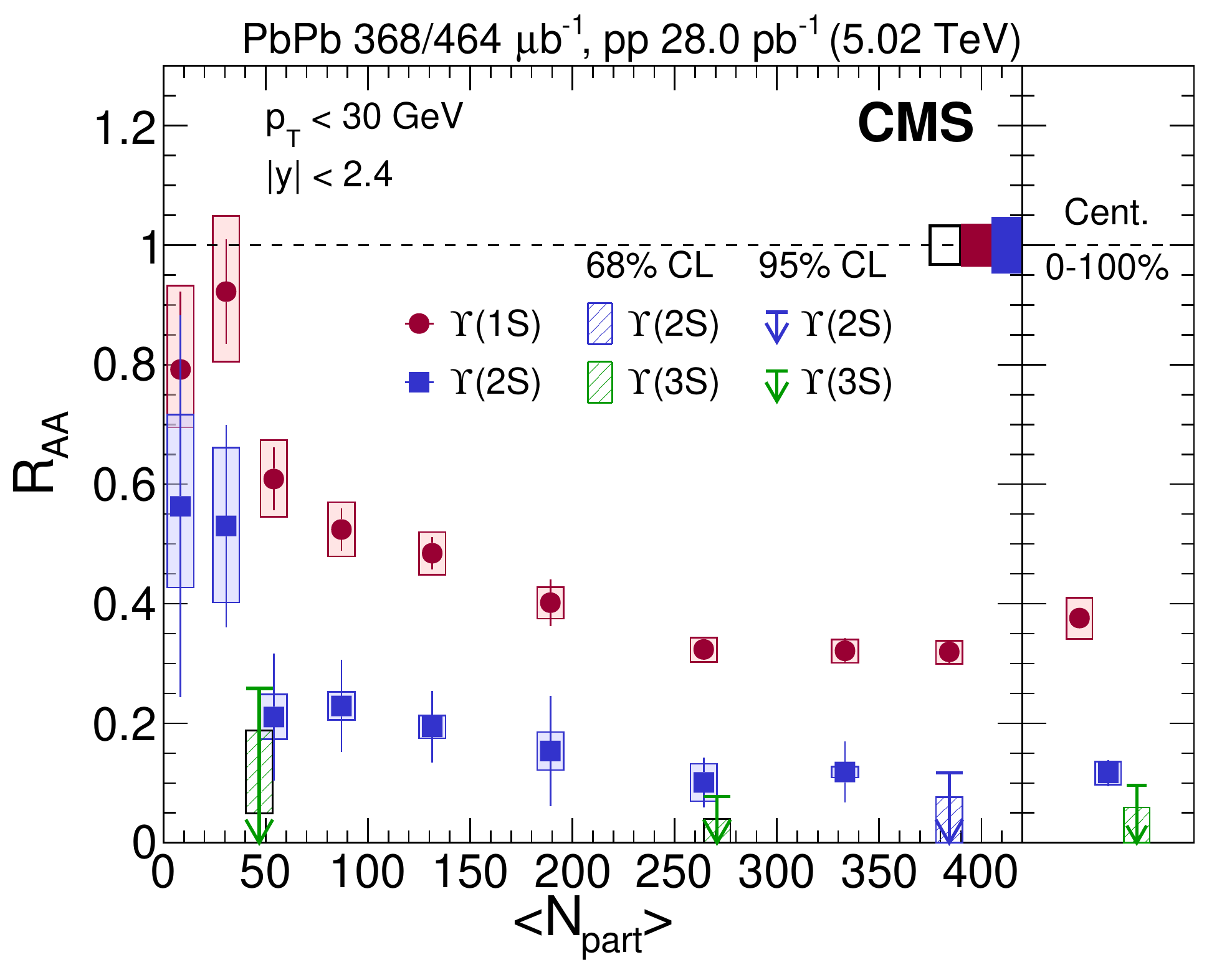}
\caption{\label{Fig:raa_cent}
Nuclear modification factors for the \PgUa, \PgUb\, and \PgUc\ mesons as a function of $\langle \Npart \rangle$. The boxes at the dashed line at unity represent global uncertainties: the open box for the integrated luminosity in \pp collisions and \NMB in \PbPb collisions, while the full boxes show the uncertainties of \pp yields for \PgUa\ and \PgUb\ states (with the larger box corresponding to the excited state). For the \PgUc\ meson, the upper limits at 68\% (green box) and 95\% (green arrow) \CL are shown.}
\end{figure}

The dependence of \RAA on \PbPb collision centrality, as quantified using the average $\langle \Npart \rangle$, is depicted in Fig.~\ref{Fig:raa_cent}. The strong suppression of the \PgUc\ meson is observed in both centrality bins studied, 0-30\% and 30-100\%. The \RAA decreases with increasing centrality in the case of the \PgUa\ and \PgUb\ mesons. A hint of this centrality dependence of \RAA for \PgUb\ was first seen in data at $\sqrtsNN = 2.76\TeV$~\cite{Khachatryan:2016xxp} and is now confirmed using the larger data sample at 5.02\TeV.

\begin{figure}[hbt!]
\centering
\includegraphics[width=0.48\textwidth]{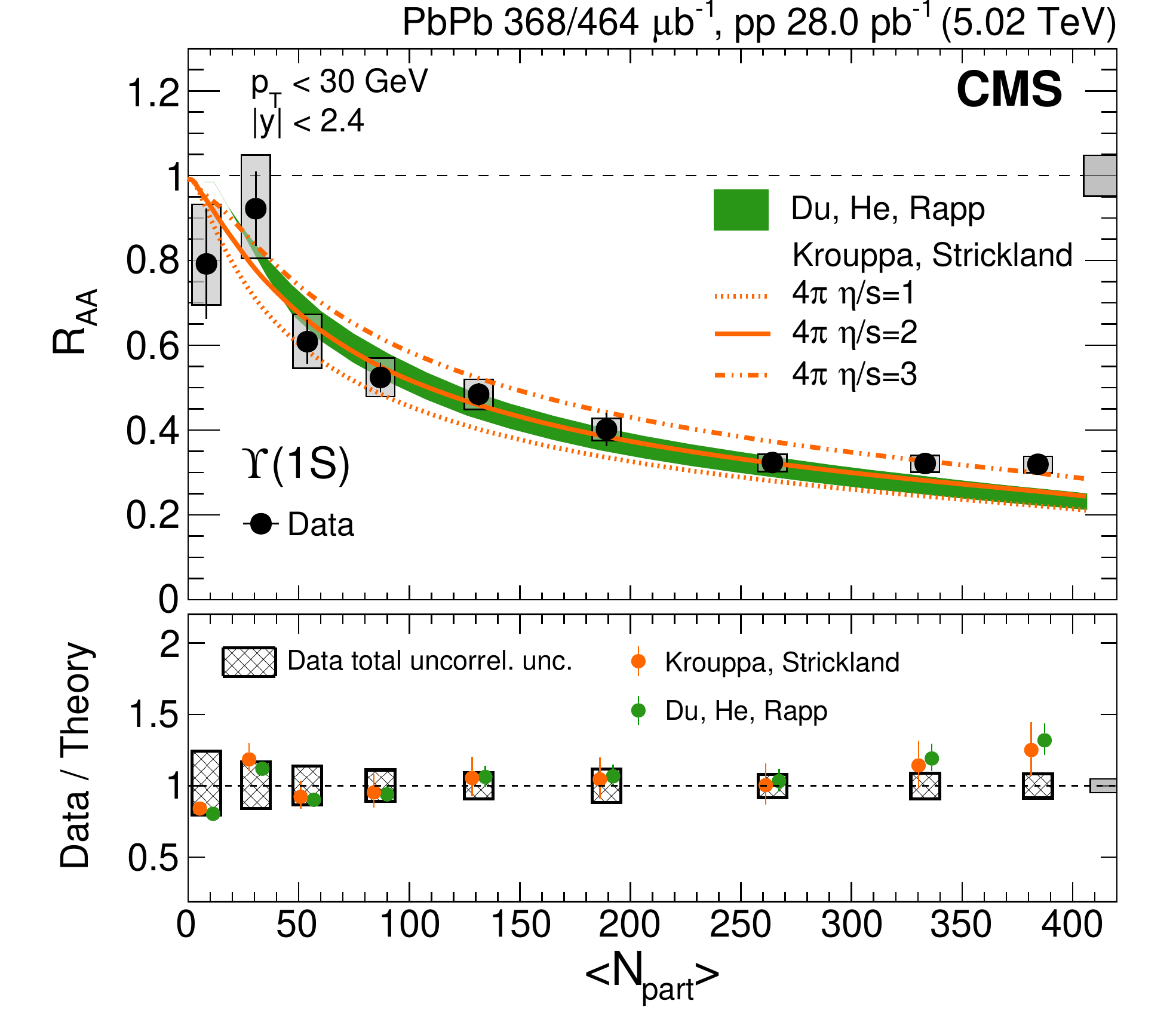}
\includegraphics[width=0.48\textwidth]{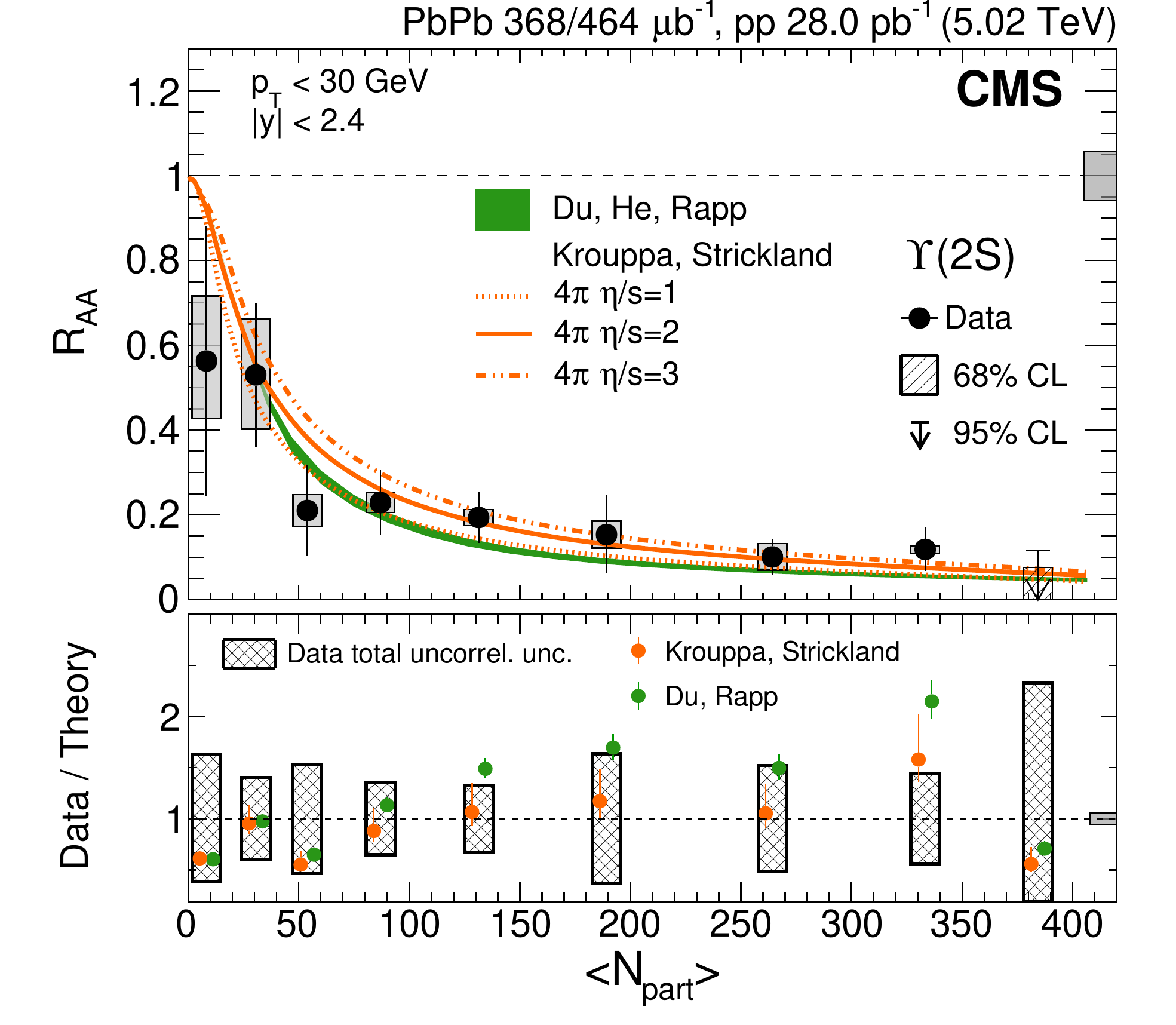}
\caption{\label{Fig:raa_th}
Nuclear modification factors for the \PgUa\ (\cmsLeft) and \PgUb\ (\cmsRight) mesons as a function of $\langle \Npart \rangle$ compared to calculations from Krouppa and Strickland~\cite{Krouppa:2016jcl},  and Du, He, and Rapp~\cite{Du:2017qkv}. The box at the dashed line at unity represents the global uncertainty from the integrated luminosity in \pp collisions,  \NMB in \PbPb collisions, and the total uncertainty in the \pp yields. The data to theory ratios are shown in the bottom panels. For Ref.~\cite{Krouppa:2016jcl}, the points correspond to the $4\pi \eta/s=2$ curve, while the error bars show the difference between this one and the other two $\eta/s$ curves. For Ref.~\cite{Du:2017qkv}, the points and error bars correspond to the center and width of the published theory band, respectively.}
\end{figure}

Figure~\ref{Fig:raa_th} shows a comparison between the measured \RAA for \PgUa\ and \PgUb\ mesons and two models of bottomonium suppression from Krouppa and Strickland~\cite{Krouppa:2016jcl}, and from Du, He, and Rapp~\cite{Du:2017qkv}. Both models incorporate color-screening effects on the bottomonium family and feed-down contributions from decays of heavier quarkonia. No regeneration in QGP or cold nuclear matter effects are considered by the first model, but are included in the second. Krouppa and Strickland treat the dynamical evolution using anisotropic hydrodynamics, where the relevant initial conditions are changed by varying the viscosity to entropy ratio, $\eta/s$, and the initial momentum-space anisotropy. The initial temperature is determined by requiring agreement with charged particle multiplicity and elliptic flow measurements. The model of Du, He, and Rapp uses a kinetic-rate equation to simulate the time evolution of bottomonium abundances in ultra-relativistic heavy ion collisions. It considers medium effects with temperature-dependent binding energies, and a lattice-QCD-based equation of state for the fireball evolution. Within the current theoretical and experimental uncertainties, both models are in agreement with the results.

\begin{figure}[hbt!]
\centering
\includegraphics[width=\cmsFigWidth]{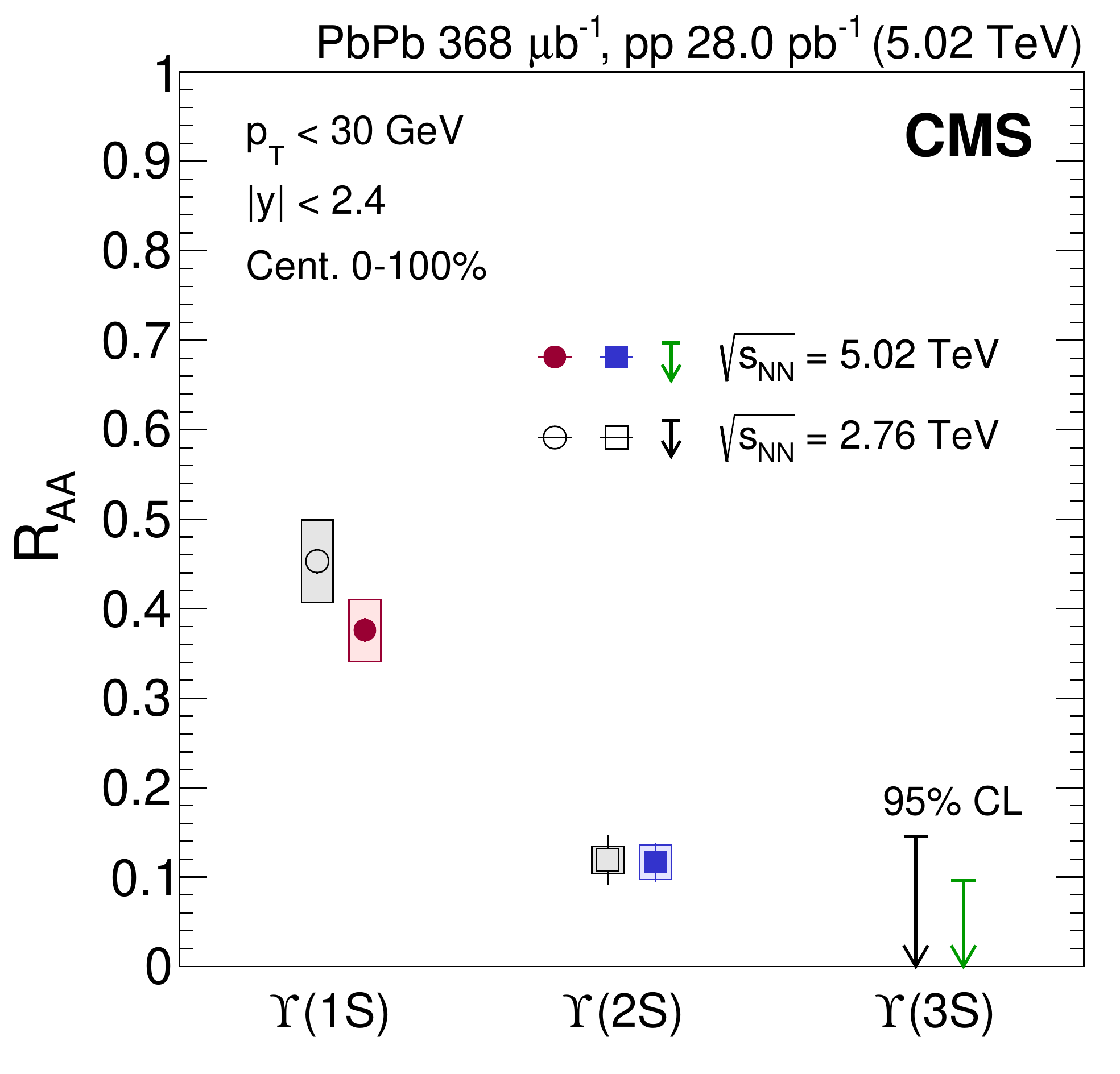}
\caption{\label{Fig:raa_comparison276}
Comparison of \RAA values for the \PgUa, \PgUb\, and \PgUc\ mesons at $\sqrtsNN = 5.02\TeV$ and $\sqrtsNN = 2.76\TeV$~\cite{Khachatryan:2016xxp} for integrated centrality in the full kinematic range. The error bars represent the statistical uncertainties and the boxes the systematic uncertainties, including global uncertainties.}
\end{figure}

Figure~\ref{Fig:raa_comparison276} compares centrality-integrated \RAA values at $\sqrtsNN = 2.76\TeV$ to those at 5.02\TeV. The centrality-integrated \RAA for \PgUa\ is measured to be $0.376 \pm 0.013 \stat \pm 0.035 \syst$, to be compared with the result at 2.76\TeV, $0.453 \pm 0.014\stat \pm 0.046\syst$~\cite{Khachatryan:2016xxp}. The suppression at 5.02\TeV is larger by a factor of $\sim 1.20 \pm 0.15$ (in which only the \TAA uncertainty was considered correlated and therefore removed),  although the two \RAA values are compatible within the uncertainties. The centrality-integrated results for the \PgUb\ and \PgUc\ states at 5.02\TeV are $\RAA(\PgUb) = 0.117\pm0.022\stat \pm 0.019\syst$ and $\RAA(\PgUc) = 0.022 \pm 0.038\stat \pm 0.016\syst$ ($<$0.096 at 95\% \CL). Despite having a bigger binding energy than the already measured \Pgy\ meson~\cite{Sirunyan:2017isk,Adam:2015isa,Khachatryan:2014bva}, no \PgUc\ meson signal is found in the \PbPb data, in any of the studied kinematic regions. This suggests a \pt- and binding-energy-dependent interplay of different phenomena affecting quarkonium states that is yet to be fully understood~\cite{Aarts:2016hap}.

Since the suppression is expected to be larger for higher temperatures in the medium, the \RAA results for the \PgUa\ meson at the two different collision energies can provide information on the medium temperature. The temperatures reported in the model of Krouppa and Strickland shown in Fig.~\ref{Fig:raa_th} are $T=641$, 631, and 629\MeV corresponding to $4\pi\eta/s=1$, 2, and 3, respectively. For the model of Du, He, and Rapp, the temperatures are in the range $T=550$--800\MeV. The models, which are also in agreement with the results at 2.76\TeV~\cite{Krouppa:2015yoa, Du:2017qkv}, predict increases in the medium temperature for \PbPb collisions of $\sim$16\% (Krouppa and Strickland) and $\sim$7\% (Du, He, and Rapp) between $\sqrtsNN = 2.76$\TeV and $\sqrtsNN = 5.02$\TeV.

\ifthenelse{\boolean{cms@external}}{}{\clearpage}

\section{Summary}
\label{sec:Summary}

Data from \pp and \PbPb collisions at $\sqrtsNN = 5.02\TeV$ collected with the CMS detector were analyzed to measure the cross sections of \PgUa, \PgUb, and \PgUc\ mesons and their nuclear modification factors as functions of \PgU\ transverse momentum (\pt) and rapidity ($y$), as well as \PbPb collision centrality. A gradual decrease in \RAA with $\langle \Npart \rangle$ for the \PgUa\ and \PgUb\ states is observed, while no significant dependence on \pt or $y$ is found in the measured region. The suppression of \PgUa\ is larger than that seen at $\sqrtsNN = 2.76\TeV$, although the two are compatible within uncertainties. The \RAA of the \PgUc\ state is measured to be below 0.096 at 95\% confidence level, making this the strongest suppression observed for a quarkonium state in heavy ion collisions to date.

\begin{acknowledgments}
The authors would like to take this opportunity to acknowledge the invaluable contributions of Roy J. Glauber to the field of heavy ion nuclear physics. Without his work, our understanding of the modifications of particle production in heavy ion collisions, of which this paper is only one of very many, would not have been possible.\\
We congratulate our colleagues in the CERN accelerator departments for the excellent performance of the LHC and thank the technical and administrative staffs at CERN and at other CMS institutes for their contributions to the success of the CMS effort. In addition, we gratefully acknowledge the computing centres and personnel of the Worldwide LHC Computing Grid for delivering so effectively the computing infrastructure essential to our analyses. Finally, we acknowledge the enduring support for the construction and operation of the LHC and the CMS detector provided by the following funding agencies: BMWFW and FWF (Austria); FNRS and FWO (Belgium); CNPq, CAPES, FAPERJ, and FAPESP (Brazil); MES (Bulgaria); CERN; CAS, MoST, and NSFC (China); COLCIENCIAS (Colombia); MSES and CSF (Croatia); RPF (Cyprus); SENESCYT (Ecuador); MoER, ERC IUT, and ERDF (Estonia); Academy of Finland, MEC, and HIP (Finland); CEA and CNRS/IN2P3 (France); BMBF, DFG, and HGF (Germany); GSRT (Greece); NKFIA (Hungary); DAE and DST (India); IPM (Iran); SFI (Ireland); INFN (Italy); MSIP and NRF (Republic of Korea); LAS (Lithuania); MOE and UM (Malaysia); BUAP, CINVESTAV, CONACYT, LNS, SEP, and UASLP-FAI (Mexico); MBIE (New Zealand); PAEC (Pakistan); MSHE and NSC (Poland); FCT (Portugal); JINR (Dubna); MON, RosAtom, RAS and RFBR (Russia); MESTD (Serbia); SEIDI, CPAN, PCTI and FEDER (Spain); Swiss Funding Agencies (Switzerland); MST (Taipei); ThEPCenter, IPST, STAR, and NSTDA (Thailand); TUBITAK and TAEK (Turkey); NASU and SFFR (Ukraine); STFC (United Kingdom); DOE and NSF (USA).

\hyphenation{Rachada-pisek} Individuals have received support from the Marie-Curie programme and the European Research Council and Horizon 2020 Grant, contract No. 675440 (European Union); the Leventis Foundation; the A. P. Sloan Foundation; the Alexander von Humboldt Foundation; the Belgian Federal Science Policy Office; the Fonds pour la Formation \`a la Recherche dans l'Industrie et dans l'Agriculture (FRIA-Belgium); the Agentschap voor Innovatie door Wetenschap en Technologie (IWT-Belgium); the Ministry of Education, Youth and Sports (MEYS) of the Czech Republic; the Council of Science and Industrial Research, India; the HOMING PLUS programme of the Foundation for Polish Science, cofinanced from European Union, Regional Development Fund, the Mobility Plus programme of the Ministry of Science and Higher Education, the National Science Center (Poland), contracts Harmonia 2014/14/M/ST2/00428, Opus 2014/13/B/ST2/02543, 2014/15/B/ST2/03998, and 2015/19/B/ST2/02861, Sonata-bis 2012/07/E/ST2/01406; the National Priorities Research Program by Qatar National Research Fund; the Programa Severo Ochoa del Principado de Asturias; the Thalis and Aristeia programmes cofinanced by EU-ESF and the Greek NSRF; the Rachadapisek Sompot Fund for Postdoctoral Fellowship, Chulalongkorn University and the Chulalongkorn Academic into Its 2nd Century Project Advancement Project (Thailand); the Welch Foundation, contract C-1845; and the Weston Havens Foundation (USA).
\end{acknowledgments}

\bibliography{auto_generated}
\ifthenelse{\boolean{cms@external}}{}{\clearpage}
\appendix
\section{Glauber model values}\label{app:Supplement}

Centrality variables computed using a Glauber model \cite{Loizides:2014vua} are summarized in Tables \ref{tab:GlauberParameters} and \ref{tab:CentralityNpart}, where $d_{\text{min}}$ is the minimum distance allowed between nucleons and $\sigma_{\mathrm{NN}}^{\text{inel}}$ is the inelastic nucleon-nucleon (NN) cross section~\cite{Khachatryan:2016odn}.

\begin{table}[htb]
\topcaption{Glauber model parameters for \PbPb collisions at $\sqrtsNN = 5.02\TeV$.}
\centering
\begin{tabular}{lc}
Parameter & Value \\  \hline
Nuclear radius (\unit{fm})                                     & $6.62 \pm 0.06$ \\
Skin depth (\unit{fm})                                           & $0.546 \pm 0.010$ \\
$d_{\text{min}}$ (\unit{fm})                                   &   $0.4 \pm 0.4$ \\
$\sigma_{\mathrm{NN}}^{\text{inel}}$ (\unit{mb}) & $70 \pm 5$ \\
\end{tabular}
\label{tab:GlauberParameters}
\end{table}

\begin{table}[htb]
\topcaption{Centrality classes, average number of participating nucleons ($\langle \Npart \rangle$), number of binary collisions (\Ncoll), and the nuclear overlap (\TAA) for \PbPb collisions at $\sqrtsNN = 5.02\TeV$, obtained using the Glauber model parameters of Table~\ref{tab:GlauberParameters}. }
\centering
\begin{tabular}{cccc}
Centrality class & $\langle \Npart \rangle$  & $\langle \Ncoll  \rangle$ & $\langle \TAA  \rangle$(\unit{mb$^{-1}$}) \\  \hline
0--5\%                &   $384.3_{-2.0}^{+1.8}$ & $1819_{-137 }^{+130 }$ & $25.98_{-0.77 }^{+0.47 }$ \\
5--10\%              &   $333.3_{-3.2}^{+3.3}$  & $1432_{-106 }^{+100 }$   & $20.46_{-0.61 }^{+0.38 }$ \\
10--20\%            &   $264.2_{-3.8 }^{+3.6 } $ & $1005_{-73 }^{+69 }$   & $14.35_{-0.46 }^{+0.33 }$ \\
20--30\%            &   $189.2_{-4.1 }^{+4.0 } $ & $606_{-44 }^{+41 }$     & $8.66_{-0.33 }^{+0.29 }$ \\
30--40\%            &   $131.4_{-4.0 }^{+4.0 } $ & $349_{-26 }^{+25 }$     & $4.98_{-0.24 }^{+0.24 }$  \\
40--50\%            &   $87.0_{-3.7 }^{+3.7 } $ &   $186_{-15 }^{+15 }$     & $2.66_{-0.17 }^{+0.18 }$  \\
50--60\%            &   $53.9_{-3.1 }^{+3.2 } $ &   $90.7_{-8.7 }^{+8.9 }$   & $1.30_{-0.12 }^{+0.12 }$  \\
60--70\%            &   $30.6_{-2.4 }^{+2.6 } $ &   $40.1_{-4.6 }^{+5.0 }$   & $0.57_{-0.064 }^{+0.071 }$  \\
70--100\%          &     $8.3_{-0.6 }^{+1.0 }$    &  $7.7_{-0.7 }^{+1.2 }$     & $0.11_{-0.011 }^{+0.018 }$ \\[\cmsTabSkip]
0--30\%              &    $270.7_{-3.4 }^{+3.2 }$ &  $1079_{-78.6 }^{+74.3 }$ & $ 15.41_{-0.47 }^{+0.33 }$\\
30--100\%          &    $46.8_{-1.2 }^{+2.4 }$   &  $98.4_{-6.4 }^{+8.0 }$  & $1.41_{-0.061 }^{+0.094 }$ \\[\cmsTabSkip]
0--100\%            &    $114_{-2.6 }^{+2.6 }$    &  $393_{-28 }^{+27 }$     & $5.61_{-0.19 }^{+0.16 }$ \\
\end{tabular}
\label{tab:CentralityNpart}
\end{table}

\cleardoublepage \section{The CMS Collaboration \label{app:collab}}\begin{sloppypar}\hyphenpenalty=5000\widowpenalty=500\clubpenalty=5000\vskip\cmsinstskip
\textbf{Yerevan Physics Institute, Yerevan, Armenia}\\*[0pt]
A.M.~Sirunyan, A.~Tumasyan
\vskip\cmsinstskip
\textbf{Institut f\"{u}r Hochenergiephysik, Wien, Austria}\\*[0pt]
W.~Adam, F.~Ambrogi, E.~Asilar, T.~Bergauer, J.~Brandstetter, E.~Brondolin, M.~Dragicevic, J.~Er\"{o}, A.~Escalante~Del~Valle, M.~Flechl, M.~Friedl, R.~Fr\"{u}hwirth\cmsAuthorMark{1}, V.M.~Ghete, J.~Grossmann, J.~Hrubec, M.~Jeitler\cmsAuthorMark{1}, A.~K\"{o}nig, N.~Krammer, I.~Kr\"{a}tschmer, D.~Liko, T.~Madlener, I.~Mikulec, E.~Pree, N.~Rad, H.~Rohringer, J.~Schieck\cmsAuthorMark{1}, R.~Sch\"{o}fbeck, M.~Spanring, D.~Spitzbart, A.~Taurok, W.~Waltenberger, J.~Wittmann, C.-E.~Wulz\cmsAuthorMark{1}, M.~Zarucki
% ...
\vskip\cmsinstskip
% ...
1:  Also at Vienna University of Technology, Vienna, Austria\\
% ...
\end{sloppypar}
\end{document}